\documentclass[12pt]{article}
\usepackage[utf8]{inputenc}
\usepackage[T1]{fontenc}
\usepackage{amsmath,amssymb,array}
\usepackage{booktabs}
\usepackage{graphics, graphicx, epsf, amsthm}
\usepackage{latexsym,amssymb}
\usepackage{color}
\usepackage{multicol}
\usepackage{hyperref}
\usepackage{subfigure}
\usepackage{fullpage}
\usepackage{natbib}

\begin{document}
\title{{\tt dbcsp}: User-friendly R package for Distance-Based Common Spacial Patterns}

\author{by Itsaso Rodr\'{i}guez, Itziar Irigoien, Basilio Sierra, and Concepción Arenas}
\maketitle

\abstract{
	Common Spacial Patterns (CSP) is a widely used method to analyse electroencephalography (EEG) data, concerning the supervised classification of brain's activity. More generally, it can be useful to distinguish between multivariate signals recorded during a time span for two different classes. CSP is based on the simultaneous diagonalization of the average covariance matrices of signals from both classes and it allows to project the data into a low-dimensional subspace. Once data are represented in a low-dimensional subspace, a classification step must be carried out. The original CSP method is based on the Euclidean distance between signals and here, we extend it so that it can be applied on any appropriate distance for data at hand. Both, the classical CSP and the new Distance-Based CSP (DB-CSP) are implemented in an R package, called {\tt dbcsp}.
}

\section{Background}

Eigenvalue and generalized eigenvalue problems are very relevant techniques in data analysis. The well-known Principal Component Analysis with the eigenvalue problem in its roots was already established by late seventies \citep{mardia79}. In mathematical terms, Common Spatial Patterns (CSP) is based on the generalized eigenvalue decomposition or the simultaneous diagonalization of two matrices to find projections in a low dimensional space. Although in algebraic terms PCA and CSP share several similarities, their main aims are different: PCA follows a non-supervised approach but CSP is a two-class supervised technique. Besides, PCA is suitable for standard quantitative data arranged in a $\mbox{`individuals} \times \mbox{variables'}$ tables, while CSP is designed to handle multivariate signals time series. That means that while for PCA each individual or unit is represented by a classical numerical vector, for CSP each individual  is represented by several signals recorded during a time span, i.e., by a $\mbox{`number of signals}\times \mbox{time span'}$ matrix. CSP allows to represent the individuals in a dimension reduced space, a crucial step given the high dimensional nature of the original data. CSP computes the average covariance matrices of signals from the two classes to yield features whose variances are optimal to discriminate the classes of measurements. Once data is projected into a low dimensional space, a classification step is carried out. The CSP technique was first proposed under the name Fukunaga-Koontz Transform in \cite{fukunaga70} as an extension of PCA and \cite{MullerGerking99} used it to discriminate electroencephalography data (EEG) in a movement task. Since then, it is a widely used technique to analyze EEG data and develop Brain Computer Interfaces (BCI), with different variations and extensions \citep{blankertz2007invariant,  blankertz07Optimizing, grosse_wentrup08, Lotte11, wang12, astigarraga16, darvish21}. \cite{samek14} offer a divergence-based framework  including several extentions of CSP. As a general term, CSP filter maximizes the variance of the filtered or projected EEG signals of one class of movements while minimizing it for the signals of the other class. Similarly, it can be used to detect epileptic activities \cite{khalid16} or other brain activities. Together with, BCI systems can be of great help to people who suffer from some disorders of cerebral palsy, or who suffer from other diseases or disabilities that prevent the normal use of their motor skills. These systems can considerably improve the quality of life of these people, for which small advances and changes imply big improvements. BCI systems can also contribute to the human vigilance detection, connected with occupations involving sustained attention tasks. Among others, CSP and variations of it have been applied to the vigilance estimation task \citep{yu19AGeneral}. 

The original CSP method is based on the Euclidean distance between signals. However, as far as we know, it was not introduced a generalization allowing the use of any  appropriate distance. The aim of the present work is  to introduce a novel Distance-Based generalization of it (DB-CSP). This generalization is of great interest, since these  techniques can also offer good solutions in other fields where multivariate time series data arise  beyond pure electroencephalography data \citep{poppe2010common,  rodriguezmoreno20}. 

Although CSP in its classical version is a very well-known technique in the field of BCI, it is not implemented in R. In addition, being DB-CSP a new extension of it, it is worth building an R package that includes both, CSP and DB-CSP techniques. The package offers functions  in a user friendly way for the less familiar users of R but it also offers complete information in its objects so that reproducible analysis can be done, as well as more advanced and customised analysis can be performed taking advantage of already well-known packages of R.

The paper is organized as follows. First, we review the mathematical formulation of the Common Spatial Patterns method. Below, the core of our contribution, we describe both the novel CSP' extension based on distances and the package {\tt dbcsp}. Then, the main functions in {\tt dbcsp} are introduced along with reproducible examples of their use. Finally, some conclusions are drawn.

\section{CSP and DB-CSP}
Let us consider we have $n$ statistical individuals or units classified in two classes $C_1$ and $C_2$, with $\#C_1 = n_1$ and $\#C_2 = n_2$. For each unit $i$ in class $C_k$  data from $c$ sources or signals are   collected during $T$ time units and therefore unit $i$ is represented in matrix $X_{ik}$ ($ i = 1, \ldots, n_k\, ; \; k=1, 2$). For instance, for electroencephalograms,
data are recorded  by a $c$-sensor cap each $t$ time units ($t=1, \ldots, T$). 
As usual, we consider that each  $X_{ik}$ is already scaled or with the appropriate pre-processing in the context of application; for instance, if working with EGG data, each signal should be band-pass filtered before its use. 

\begin{figure}
	\centering
	\includegraphics[width=13cm]{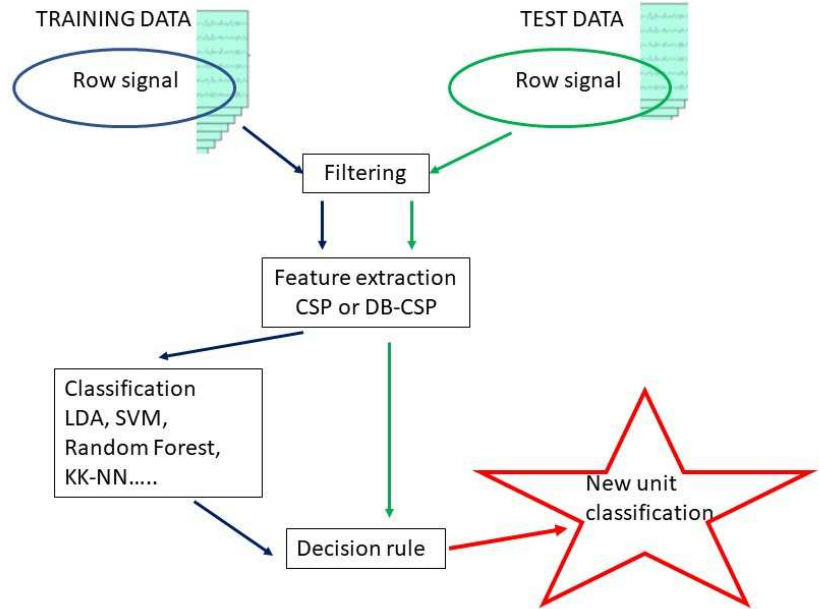}
	\caption{The flow-chart shows how the consecutive steps of filtering, feature extraction and the construction of a classification model can be used to classify a new data. }
	\label{fig:F1}
\end{figure}

The goal is to 
classify a new unit $X$ in $C_1$ or $C_2$. To this end, first a projection into a low-dimensional subspace is carried out. Then, classically the Linear Discriminant classifier (LDA) is applied taking as input data for the classifier the log-variance of the projections obtained in the first step. It is obvious that the importance of the technique lies mainly in the first step, and once it is done, LDA or any other classifiers could be applied. Based on that, we focus on how this projection into a low-dimensional space is done, from the classical CSP point of view as well as its novel extension DB-CSP (see Figure \ref{fig:F1}).

\subsubsection{Classical CSP}
The main idea is to use a linear transform to project or filter data into low-dimensional subspace with a projection matrix, in such a way that each row consists of weights for signals. This transformation maximizes the variance of two-class signal matrices. The method performs a simultaneous diagonalization of the covariance matrices of both classes.
Given data $X_{11}, \ldots, X_{n_1 1}$ (matrices $c \times T$) from class $C_1$ and $X_{12}, \ldots, X_{n_2 2}$ (also matrices $c \times T$) from class $C_2$, the following steps are needed:
\begin{itemize}
	\item All matrices are standardized so that traces of $X_{i k}X_{i k}'$ are the same.
	\item Compute average covariance matrices:
	$$B_k = \frac{1}{n_k}\sum_{i=1}^{n_k}X_{i k} X_{i k}' \, , \quad k=1,2$$ 
	\item Look for directions $W= (\mathbf{w}_1, \ldots, \mathbf{w}_c) \in \mathbb{R}^{c \times c}$ according to the criterion:
	
	\begin{align*} 
	\mbox{ Maximize}\; & tr(W'B_1W)\\
	\mbox{ subject to}\;  & W'(B_1 + B_2)W = I\\
	\end{align*}
	
	The solution is given by the generalized spectral decomposition
	$B_1\mathbf{w} = \lambda B_2 \mathbf{w}$
	choosing the  first and the last $q$ eigenvectors: $W_{CSP}= (\mathbf{w}_1, \ldots, \mathbf{w}_q, \mathbf{w}_{c-q+1}, \ldots, \mathbf{w}_c)$.
\end{itemize}

Vectors $\mathbf{w}_j$ offer weights so that new signals $X_{i 1}'\mathbf{w}_j$  and $X_{i 2}'\mathbf{w}_j$ have big and low variability for the first $q$ vectors ($j=1, \ldots, q$) respectively, and vice-versa for the last $q$ vectors ($j=c-q+1, \ldots, c$). To clarify the notation and interpretation, let us denote $\mathbf{a}_j=\mathbf{w}_j$ the first $q$ vectors and $\mathbf{b}_j=\mathbf{w}_{c+1-j}$ the last $q$. That way, and broadly speaking,  variability of elements in $C_1$ is big when projecting on vectors $\mathbf{a}_j$ and low on vectors $\mathbf{b}_j$, and vice-versa, for elements in class $C_2$.

Finally, the log-variability of these new and few $2q$ signals are considered as input for the classification, which classically is the Linear Discriminant Analysis. Obviously, any other classification technique can be used.

\subsubsection{Distance-based CSP}
Following the commented ideas, the Distance-Based CSP (DB-CSP) is an extension of the classical CSP method. In the same way as the classical CSP, DB-CSP gives some weights to the original sources or signals and obtains new and few $2q$ signals useful for the discrimination between the two classes. Nevertheless, the considered distance between the signals sources can be any other than the Euclidean. The steps are the following:

\begin{itemize}
	\item  Compute an appropriate distance measure between sources and the double-centered inner product:
	\begin{align*} 
	X_{i k} & \rightarrow D_{i k} \rightarrow  P_{ i k}=-1/2HD_{i k}^{(2)}H \, , \quad i=1,\ldots, n_k; \; k=1,2
	\end{align*}
	where $H$ stands for the centering matrix and the superindex in brackets $(2)$ for squared elements in the matrix. Again, all matrices are standardized so that all traces of $X_{ik}X_{ik}'$ are the same.

	\item Compute average  distance-based covariance matrices:
	$$B_k^* = \frac{1}{n_k}\sum_{i=1}^{n_k}\left(P_{ i k}P_{ i k}' + X_{i k}\mathbf{x}_{i k}\mathbf{1}' + \mathbf{1}\mathbf{x}_{i,k}'X_{i k}' - \mathbf{x}_{i k}'\mathbf{x}_{i k}\mathbf{1}\mathbf{1}' \right)$$
	
	where $\mathbf{x}_{i k} = \frac{1}{c}\mathbf{1}'\mathbf{X}_{i k}$, and $k=1, 2$.
\end{itemize}

Once we have the covariance matrices related to the chosen distance matrix, the directions are found as in classical CSP and new signals $X_{i k}'\mathbf{a}_j$, $X_{i k}'\mathbf{b}_j$ are built ($j=1, \ldots, q$). Again, for individuals in class $C_1$ the projections on vectors $\mathbf{a}$ and $\mathbf{b}$ are big and low respectively; for individuals in class $C_2$ the other way round.

\bigskip

It is important to note that if the chosen distance does not produce a positive definite covariance matrix, it must be replaced by a similar one that is positive definite.

\bigskip

When the selected distance is the Euclidean, then, {\bf DB-CSP reduces to classical CSP}.

Once the $q$ directions $\mathbf{a}_j$ and $\mathbf{b}_j$ are calculated, new $2q$ signals are built. Many interesting characteristics of the new signals could be extracted, although the most important in the procedure is the variance. Those characteristics of the new signals are the input data for the classification step.

\section{Implementation}
In this section, the structure of the package and the functions implemented are explained. The {\tt dbcsp} package was developed for the free statistical R environment (http://www.r-project.org) and it is  available from the Comprehensive R Archive Network (CRAN) at \url{https://cran.r-project.org/web/packages/dbcsp/index.html}.

\subsection{Input}
The input data are the corresponding $n_1$ and $n_2$ matrices $X_{i k}$  of the $n$ units classified in classes $C_1$ and $C_2$, respectively ($i=1, \ldots, n_k$\, ; \; $k=1, 2$).   Let {\tt x1} and {\tt x2} be  two lists of length $n_1$ and $n_2$, respectively, with $X_{i k}$  matrices ($c \times T$) as elements of the lists.  We also consider that new items to be classified are in list {\tt xt}.
The aforementioned first step of the method is carried out by building the object called {\tt "dbcsp"}.

\subsection{{\tt dbcsp} object}

The {\tt dbcsp} object is an S4 class created to compute the projection vectors $W$. The object has the following slots:\\

\begin{itemize}
	\item{\bf Slots}
	
	{\tt X1 = "list"}, {\tt X2 = "list"}, the lists {\tt X1} and {\tt X2} (lengths $n_1$ and $n_2$) containing the matrices $X_{i k}$ for the two classes $C_1$ and $C_2$, respectively ($i=1, \ldots, n_k \, ; \; k=1,2$).\\ 
	
	{\tt q = "integer"},  to determine the number of pairs of eigenvectors $\mathbf{a}_j$ and $\mathbf{b}_j$ that are kept. By default {\tt q =15}. \\
	
	{\tt labels = "character"}, vector of two strings indicating labels names, by default names of elements in {\tt X1} and {\tt X2}.\\
	
	{\tt type = "character"}, to set the type of distance to be considered, by default {\tt type='EUCL'}. The supported distances are these ones:
	\begin{itemize}
		\item Included in  {\textbf{\tt TSdist}}: {\tt infnorm, ccor, sts,...
		}
		\item Included in  {\textbf{\tt parallelDist}}: {\tt bhattacharyya, bray,... 
		}
	\end{itemize} 
	
	{\tt mixture = "logical"}, logical value indicating whether to use mixture of distances or not (EUCL + other), by default {\tt mixture=FALSE}.\\
	
	{\tt w = "numeric"}, weight for the mixture of distances  $D_{\mbox{mixture}} = w D_{\mbox{euclidea}} + (1-w) D_{\mbox{type}}$, by default {\tt w=0.5}.\\
	
	{\tt training = "logical"}, logical value indicating whether to perform the classification or not. If {\tt classification = TRUE}, LDA discrimination based on the log-variances of the projected sources is considered, following the classical approach in CSP. \\
	
	{\tt fold = "integer"}, integer value, by default {\tt fold=10}. It controls the number of partitions for the $k$-fold validation procedure, if the classification is done. \\
	
	{\tt seed = "numeric"}, numeric value, by default {\tt seed=NULL}. Set a seed in case you want to be able to replicate the results.\\
	
	{\tt eig.tol= "numeric"}, numeric value, by default {\tt eig.tol=1e-06}. If the minimum eigenvalue is below this tolerance, average covariance matrices are replaced by the most similar matrix that is definite positive. It is done via function {\tt nearPD} in  {\textbf{\tt Matrix}}.\\
	
	{\tt out= "list"}, list containing elements of the output. Mainly, matrix $W$ with vectors $\mathbf{a}_j$ and $\mathbf{b}_j$ in element {\tt vectors}, log-variances of filtered signals in  {\tt proy} and partitions considered in the $k$-fold approach with reproducibility purposes.
	
	\bigskip
	
	\item{\bf Usage}\\
	Following the standard procedure in {\tt R}, an instance of a class {\tt dbcsp} is created via the {\tt new()} constructor function:\\
	
	{\tt new("dbcsp", X1 = x1, X1 = x2)}\\
	
	Slots {\tt X1} and {\tt X2} are compulsory since they contain the original data. When a slot is not specified, the default value is considered. First, the S4 object of class {\tt dbcsp} must be created.  By default, the Euclidean distance is used, nevertheless it can be changed. For instance, "Dynamic Transform Distance" \citep{giorgino09} can be set:\\
	
	{\tt mydbcsp <- new('dbcsp', X1=x1, X2=x2, type='dtw')}\\
	
	or a mixture between this distance and the euclidean can be indicated by:\\
	
	{\tt mydbcsp.mix <- new('dbcsp', X1=x1, X2=x2, labels=c("C1", "C2"), mixture=TRUE, w=0.4,type="dtw")}\\
	
	The object contains all the information to carry out classification task in a lower dimension space. 
\end{itemize}

\subsubsection{Functions {\tt plot} and {\tt boxplot}}
For exploratory and descriptive purposes, the original signals $X_{i k}$ and the projected ones can be plotted for selected individual $i$ in class $k$, and selected pair of dimensions $\mathbf{a}_j$ and $\mathbf{b}_j$ ($i= 1, \ldots, n_k$, $k=1,2$).

\begin{itemize}
	\item{\bf Usage}
	
	{\tt plot(mydbcsp)}\\
	
	\item{\bf Arguments}
	
	{\tt object}, an object of class {\tt dbcsp}\\
	
	{\tt class}, integer to indicate which of both classes to access (1 or 2), by default {\tt class=1}.\\
	
	{\tt  index}, integer to indicate which instance of the class to plot, by default {\tt index=1}.\\
	
	{\tt vectors}, integer to indicate which $j$ projected signals are to be plotted. By default all the vectors used in the projection are plotted.\\
	
	{\tt pairs} logical, if {\tt TRUE} the pairs $\mathbf{a}_j$ and $\mathbf{b}_j$ of the indicated indices are also shown, by default {\tt pairs=TRUE}.\\
	
	{\tt  before} logical, if {\tt TRUE} the original signals are plotted, by default {\tt before=TRUE}.\\
	
	{\tt  after}	logical, if {\tt TRUE} the signals after projection are plotted, by default {\tt after=TRUE}.\\
	
	{\tt  legend}	logical, if {\tt TRUE}, a legend for filtered signals is shown,  by default {\tt legend=FALSE}.\\
\end{itemize}

\bigskip
Besides, the log-variances of the projected signals of both classes can be shown in boxplots. This graphic can help to understand the discriminative power that is in the low-dimension space.

\begin{itemize}
	\item{\bf Usage}
	
	{\tt boxplot(mydbcsp)}\\
	
	\item{\bf Arguments}
	
	{\tt object}, an object of class {\tt dbcsp}\\
	
	{\tt vectors}, integer or vector of integers, indicating the index of the projected vectors to plot, by default {\tt index=1}.\\
	
	{\tt pairs} logical, if {\tt TRUE} the pairs $\mathbf{a}_j$ and $\mathbf{b}_j$ of the indicated indices are also shown, by default {\tt pairs=TRUE}.\\
	
	{\tt show\_log} logical, if {\tt TRUE} the logarithms of the variances are displayed, else the variances, by default {\tt show\_log=TRUE}.\\
\end{itemize}

\bigskip

It is worthy to take into account that in the aforementioned functions, values in argument {\tt vectors}  must lie between 1 and $2q$, being $q$ the number of dimensions used to perform the DB-CSP algorithm when creating the {\tt dbcsp} object. Therefore, values 1 to $q$ correspond to vectors $\mathbf{a}_1$ to $\mathbf{a}_q$ and values $q+1$ to $2q$ correspond to vectors $\mathbf{b}_1$ to $\mathbf{b}_q$. Then, if {\tt pairs=TRUE}, it is recommended that values in argument {\tt vectors} are in $\{1, \ldots, q\}$, since their pairs are plotted as well. In case that values are above $q$, it should be noted that they correspond to vectors $\mathbf{b}_1$ to $\mathbf{b}_q$. For instance,  if {\tt q=15} and {\tt boxplot(object, vectors=16, pairs=FALSE)}, vector $\mathbf{b}_1$ $(16-q=1)$  is shown.

\subsubsection{Function {\tt selectQ}, Function {\tt train}  and Function {\tt predict}}

The functions in this section help the classification step in the procedure. Function {\tt selectQ} helps to find an appropriate dimension needed for the classification. Given different values of dimensions, the accuracy related to each dimension is calculated so that the user can assess which dimension of the reduced space can be sufficient. A $k$-fold cross-validation approach or a holdout approach can be followed.  Function {\tt train} performs the Linear Discriminant classification based on the log-variances of the dimensions built in the {\tt dbcsp} object. The accuracy of the classifier is computed based on $k$-fold validation procedure.   Finally,  function {\tt predict} performs the classification of new individuals. 

\begin{itemize}
	\item{\bf Usage of {\tt selectQ}} \\
	{\tt selectQ(mydbcsp)} \\
	
	\item{\bf Arguments}
	
	{\tt object}, an object of class {\tt dbcsp}\\
	
	{\tt Q}, vector of integers which represents the dimensions to use, by default {\tt Q=c(1,2,3,5,10,15)}.\\
	
	{\tt train\_size}, float between 0.0 and 1.0 representing the proportion of the data set to include in the train split, by default {\tt train\_size=0.75}.\\
	
	{\tt CV}, logical indicating whether a $k$-fold cross validation must be performed or a hold-out approach (if {\tt TRUE}, {\tt train\_size} is not used), by default {\tt CV=FALSE}.\\
	
	{\tt folds} integer, number of folds to use if {\tt CV} is performed.\\
	
	{\tt seed} numeric value, by default {\tt seed=NULL}. Set a seed in case you want to be able to replicate the results.\\
	
\end{itemize}

This  functions returns the accuracies related to each dimension set in {\tt Q}. If {\tt CV=TRUE}, the mean accuracy as well as the standard deviation among folds is also returned.

\medskip

\begin{itemize}
	\item{\bf Usage of {\tt train}}
	
	{\tt train(mydbcsp)} or embedded as a parameter in:\\ {\tt new('dbcsp', X1=x1, X2=x2, training=TRUE, type="dtw")}\\
	
	\item{\bf Arguments}
	
	{\tt object}, an object of class {\tt dbcsp}\\
	
	{\tt selected\_q}, integer value indicating the number of vectors to use when training the model. By default all dimensions considered when creating the object {\tt dbcsp}.
	
	\medskip
	
	Besides, arguments {\tt seed} and {\tt fold} are available.
\end{itemize}

It is important to note that in this way a classical analysis can be carried out, in the sense of:
\begin{itemize}
	\item LDA is applied based on the log-variances of the dimensions indicated by the user in, {\tt select\_q}; 
	\item percentage of good classification is obtained via $k$-fold cross validation.
\end{itemize}
However, it is evident that it may be of interest to use other classifiers or other characteristics in addition to or different from log-variances. This more advanced procedure is explained below. See the basic analysis of the "User guide with real example" section in order to visualize and follow the process of a first basic/classic analysis.\\

\begin{itemize}
	\item{\bf Usage of {\tt predict}} \\
	{\tt predict(mydbcsp, X\_test=xt)} \\
	
	\item{\bf Arguments}
	
	{\tt object}, an object of class {\tt dbcsp}\\
	
	{\tt X\_test}, list of matrices for be classified.\\
	
	{\tt true\_targets}, optional, if available, vector of true labels of the instances. Note that they must match the name of the labels used when training the model.
\end{itemize}

\section{User guide with a real example}
To show an example beyond pure electroencephalography data, an Action Recognition data is considered. Besides, in order to have a reproducible example to show the use of the implemented functions and the results they offer, this Action Recognition data set is included in the package. The data set contains the skeleton data extracted from videos of people performing six different actions, recorded by a semi-humanoid robot. It consists of a total of 272 videos with 6 action categories. There are around 45 clips in each category, performed by 46 different people. Each instance is composed of 50 signals ($xy$ coordinates for 25 body key points extracted using OpenPose \citep{cao2019openpose}), where each signal has 92 values, one per frame. These are the six categories included in the data set:
\begin{enumerate}
	\item Come: gesture for telling the robot to come to you. There are 46 instances for this class.
	\item Five: gesture of `high five'. There are 45 instances for this class.
	\item Handshake: gesture of handshaking with the robot. There are 45 instances for this class.
	\item Hello: gesture for telling hello to the robot. There are 44 instances for this class.
	\item Ignore: ignore the robot, pass by. There are 46 instances for this class.
	\item Look at: stare at the robot in front of it. There are 46 instances for this class.
\end{enumerate}

The data set is accessible via {\tt AR.data}. Each class is a list of matrices of $[K \times num\_frames]$ dimensions, where $K=50$ signals and $num\_frames=92$ values. As mentioned before, the 50 signals represent the $xy$ coordinates of 25 body key points extracted by OpenPose. 

For example, two different classes can be accessed this way:
\begin{verbatim}
x1 <- AR.data$come
x2 <- AR.data$five
\end{verbatim}
where, {\tt x1} is a list of 46 instances of $[50x92]$ matrices and {\tt x2} is a list of 45 instances of $[50x92]$ matrices. First a basic/classic analysis is performed.

\subsection{Basic/classic analysis}

Suppose an analysis using 15-dimensional projections and the Euclidean distance. At a first step the user can obtain  vectors $W$ by:

\begin{verbatim}
x1 <- AR.data$come
x2 <- AR.data$five
mydbcsp <- new('dbcsp', X1=x1, X2=x2, q=15, labels=c("C1", "C2"))
summary(mydbcsp)
\end{verbatim}

Creating the object {\tt mydbcsp}, the vectors $W$ are calculated. As indicated in parameter {\tt q=15}, the first and last 15 eigenvectors are retained. With {\tt summary}, the obtained output is:

\begin{verbatim}
There are 46 instances of class C1 with [50x92] dimension.
There are 45 instances of class C2 with [50x92] dimension.
The DB-CSP method has used 15 vectors for the projection.
EUCL distance has been used.
Training has not been performed yet.
\end{verbatim}

Now, if the user knows from the beginning that 3 is an appropriate dimension, the classification step could be done while creating the object.  Using classical analysis, with for instance 10-fold, LDA as classifier and log-variances as characteristics, the corresponding input and summary output are:

\begin{verbatim}
mydbcsp <- new('dbcsp', X1=x1, X2=x2, q=3, labels=c("C1", "C2"), training=TRUE, fold = 10, seed = 19)
summary(mydbcsp)
\end{verbatim}

\begin{verbatim}
There are 46 instances of class C1 with [50x92] dimension.
There are 45 instances of class C2 with [50x92] dimension.
The DB-CSP method has used 3 vectors for the projection.
EUCL distance has been used.
An accuracy of 0.9130556 has been obtained with 10 fold cross validation and using 3 vectors when training.
\end{verbatim}

If a closer view of the accuracies  among the folds are needed, the user can get them from the {\tt out} slot of the object:

\begin{verbatim}
# Accuracy in each fold
mydbcsp@out$folds_acc

# Intances belonging to each fold
mydbcsp@out$used_folds
\end{verbatim}

\subsection{Basic/classic analysis selecting the value of $q$}

Furthermore, it is clear that the optimal value of $q$ should be chosen based on the percentages of good classification. It is worth mentioning that the LDA is applied on the $2q$ projections, as set in the object building step. It is interesting to measure how many dimensions would be enough using {\tt selectQ} function:

\begin{verbatim}
mydbcsp <- new('dbcsp', X1=x1, X2=x2, labels=c("C1", "C2"))
selectDim <- selectQ(mydbcsp, seed=30, CV=TRUE, fold = 10) 
\end{verbatim}

\begin{verbatim}
selectDim
Q       acc         sd
1  1 0.7663889 0.12607868
2  2 0.9033333 0.09428818
3  3 0.8686111 0.11314534
4  5 0.8750000 0.13289537
5 10 0.8797222 0.09513230
6 15 0.8250000 0.05257433
\end{verbatim}

Since the $10$-fold cross-validation approach is chosen, the mean accuracies as well as the corresponding standard deviations are returned. Thus, with Linear Discriminant Analysis (LDA), log-variances as characteristics, it seems that dimensions related to first and last $q=2$ eigenvectors ($2\times 2$ dimensions in total) are enough to obtain a good classification, with an accuracy of 90\%. Nevertheless, it is also observed that variation among folds can be relevant.

\begin{figure}
	\centering
	\includegraphics{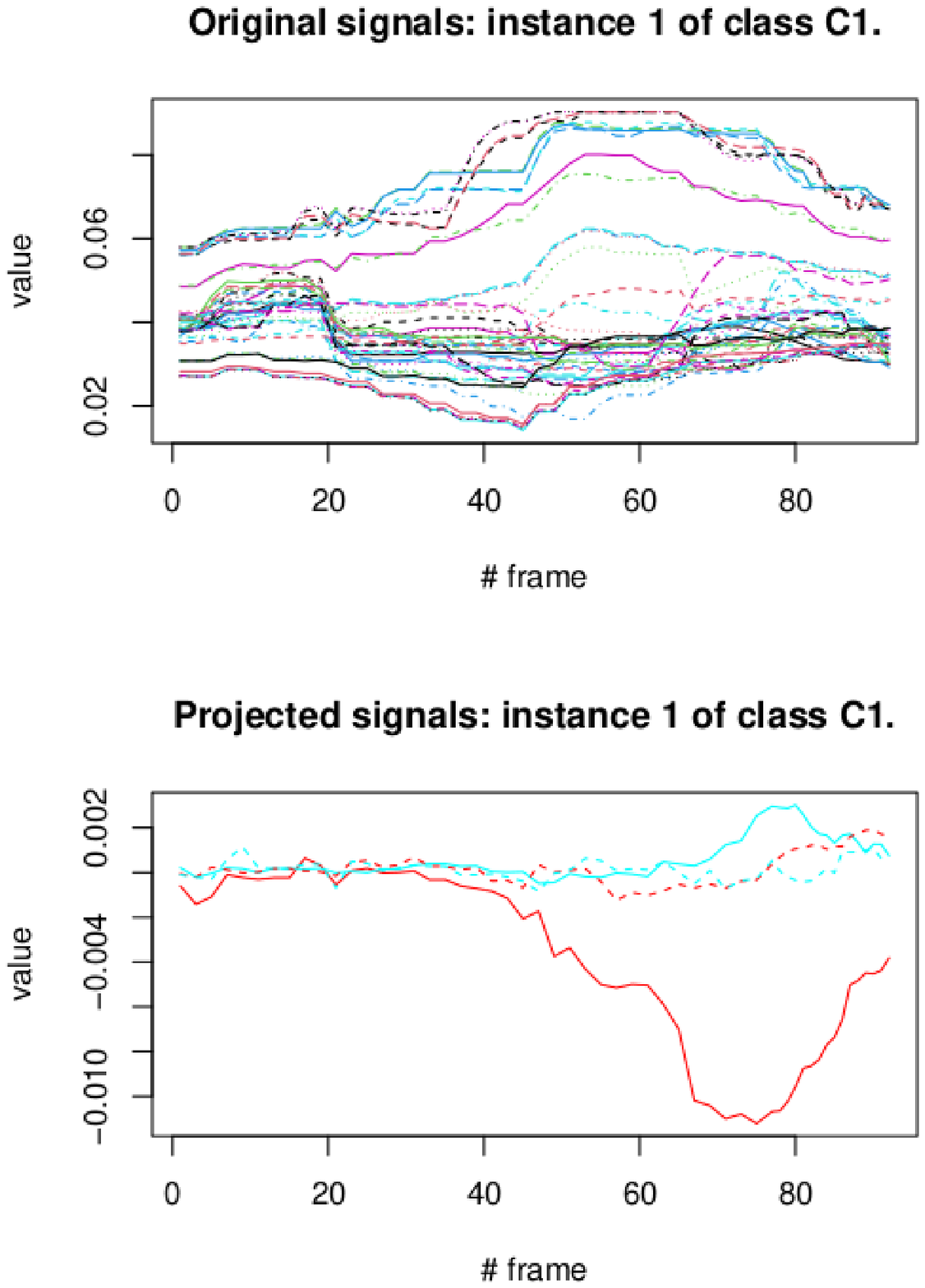}
	\caption{First video of class $C_1$ represented as its original version (top) and as the projections on vectors $\mathbf{a}_1$ and $\mathbf{a}_2$ (continuous lines) and $\mathbf{b}_1$ and $\mathbf{b}_2$ (dotted lines). }
	\label{fig:plot1}
\end{figure}

To visualize what is the representation in the reduced dimension space function {\tt plot} can be used. For instance, to visualize the first unit of the first class, based on projections along the 2 first and last vectors ($\mathbf{a}_1,  \mathbf{a}_2$ and $\mathbf{b}_1, \mathbf{b}_2$):

\begin{verbatim}
plot(mydbcsp, index=1, class=1, vectors=1:2)
\end{verbatim}
In the top  graphic of Figure \ref{fig:plot1} it can be seen what is the representation of the first video of class $C_1$ given by matrix  $X_{11}$, where the horizontal axis represents the frames of the video and the lines are the positions of the body key points (50 lines). In the bottom graphic, the same video is represented in a reduced space where the video is represented by the new signals (only 4 lines).

To have a better insight of the discriminating power of the new signals in the reduced dimension space, we can plot the corresponding log-variances of the new signals. Parameter {\tt vectors} in function  {\tt boxplot} sets which are the eigenvectors considered to plot.

\begin{verbatim}
boxplot(mydbcsp, vectors=1:2)
\end{verbatim}

\begin{figure}
	\centering
	\includegraphics{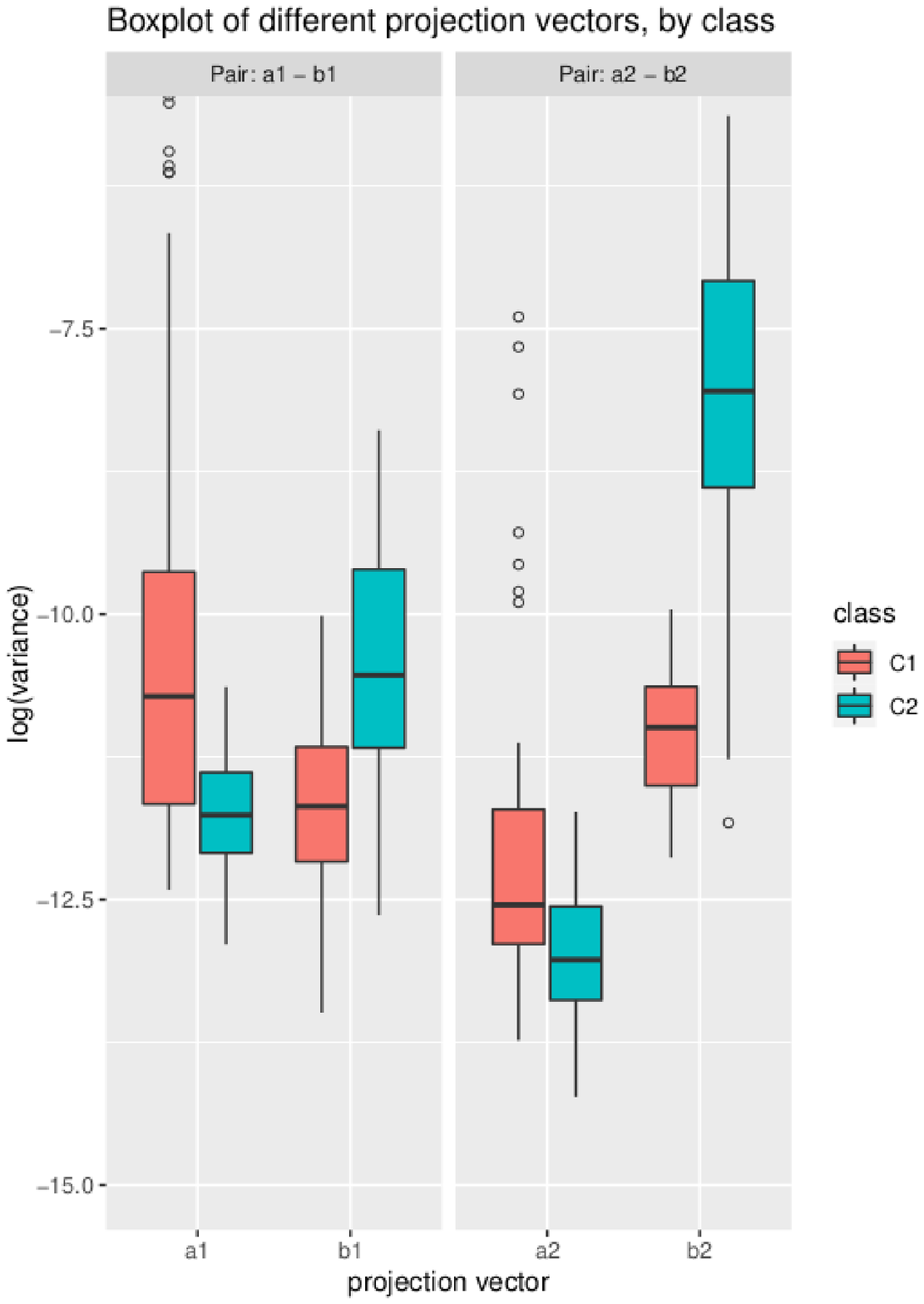}
	\caption{Log-variabilities of the projected signals on vectors $\mathbf{a}_1$ and $\mathbf{a}_2$ and $\mathbf{b}_1$ and $\mathbf{b}_2$ and separated by classes $C_1$ and $C_2$. }
	\label{fig:boxplot1}
\end{figure}
In Figure \ref{fig:boxplot1} it can be seen that variability of projections on the first eigenvector direction ($log(VAR(X_{i k}'\mathbf{a}_1))$) are big for elements in {\tt x1}, but small for elements in {\tt x2}.  Analogously, projecting on the last dimension ($log(VAR(X_{i k}'\mathbf{b}_1))$), low variability is hold in {\tt x1} and big variability in {\tt x2}. The same pattern holds when projecting on vectors $\mathbf{a}_2$ and $\mathbf{b}_2$.\\




\subsection{Basic/classic analysis new unit classification}

Once the value of $q$ has been decided and the accuracy of the classification is known, the classifier should be built (through {\tt train()}) so that the user can proceed to predict  the class a new action  hold in a video belongs to, using the function {\tt predict}. For instance, with only illustrative purpose, we can classify the first 5 videos which are stored in {\tt x1}. 

\begin{verbatim}
mydbcsp <- train(mydbcsp, selected_q=2, verbose=FALSE)
xtest <- x1[1:5]
outpred <- predict(mydbcsp, X_test=xtest)
\end{verbatim}

In case it is known the label of the testing items, the latter function returns the accuracy.

\begin{verbatim}
outpred <- predict(mydbcsp, X_test=xtest, true_targets= rep("C1", 5))
\end{verbatim}

Finally notice that the user could use any other distance instead of the Euclidean between the signals to compute the important directions $\mathbf{a}_j$ and $\mathbf{b}_j$. For instance, in this case it could be appropriate to use the Dynamic Time Warping distance, setting so in the argument {\tt type="dtw"}:
\begin{verbatim}
# Distance DTW
mydbcsp.dtw <- new('dbcsp', X1=x1, X2=x2, labels=c("C1", "C2"), type="dtw")
\end{verbatim}

\section{Extending the example}

In the previous section a basic workflow to use functions implemented in {\tt dbcsp} is presented. Nevertheless, it is straightforward to extend the procedure. Once the interesting directions in $W$ are calculated through {\tt dbcsp}, other summarizing characteristics beyond the variance could be extracted from the projected signals, as well as other classifiers could be used in the classification step. Here it is shown how once the eigenvectors are extracted from an object {\tt dbcsp}, several characteristics could be extracted from the signals and a new {\tt data.frame} can be built so that any other classification technique could be applied. In this example we worked with {\tt caret} package to apply different classifiers.
It is important to pay attention to which the train and test sets are, so that the vectors are computed based only on training set instances.

\begin{verbatim}
# Establish training and test data
n1 <- length(x1)
trainind1 <- rep(TRUE, n1)
n2 <- length(x2)
trainind2 <- rep(TRUE, n2)
set.seed(19)
trainind1[sample(1:n1, 10, replace=FALSE)] <- FALSE
trainind2[sample(1:n2, 10, replace=FALSE)] <- FALSE
x1train <- x1[trainind1]
x2train <- x2[trainind2]

# Extract the interesting directions
vectors <- new('dbcsp', X1=x1train, X2=x2train, q=5, labels=c("C1", "C2"))@out$vectors

# Function to calculate the desired characteristics from signals
calc_info <- function(proj_X, type){
values <- switch(type,
'var' = values <-  plyr::laply(proj_X, function(x){apply(x,1,var)}),
'max' = values <-  plyr::laply(proj_X, function(x){apply(x,1,max)}),
'min' = values <- plyr::laply(proj_X, function(x){apply(x,1,min)}),
'iqr' = values <- plyr::laply(proj_X, function(x){
apply(x,1,function(y){
q <- quantile(y, probs = c(0.25, 0.75))
q[2] -q[1]
})
})
)
return(values)
}
\end{verbatim}
By means of  this latter function, besides the variance of the new signals, the maximum, the minimum, and the interquartile range can be extracted.

Then, imagine we want to perform our classification step with the interquartile range information besides the log-variance. 

\begin{verbatim}
# Project units of class C1 and 
projected_x1 <- plyr::llply(x1, function(x,W) t(W)%*%x, W=vectors)

# Extract the characteristics
logvar_x1 <- log(calc_info(projected_x1,'var'))
iqr_x1 <- calc_info(projected_x1,'iqr')
new_x1 <- data.frame(logvar=logvar_x1, iqr=iqr_x1)

# Similarly for units of class C2
projected_x2 <- plyr::llply(x2, function(x,W) t(W)%*%x, W=vectors)
logvar_x2 <- log(calc_info(projected_x2,'var'))
iqr_x2 <- calc_info(projected_x2,'iqr')
new_x2 <- data.frame(logvar=logvar_x2, iqr=iqr_x2)


# Create dataset for classification
labels <- rep(c('C1','C2'), times=c(n1,n2))
new_data <- rbind(new_x1,new_x2)
new_data$label <- factor(labels)
new_data_train <- new_data[c(trainind1, trainind2), ]
new_data_test <- new_data[!c(trainind1, trainind2), ]

# Random forest
trControl <- caret::trainControl(method = "none")
rf_default <- caret::train(label~.,
data = new_data_train,
method = "rf",
metric = "Accuracy",
trControl = trControl)
rf_default

# K-NN
knn_default <- caret::train(label~.,
data = new_data_train,
method = "knn",
metric = "Accuracy",
trControl = trControl)
knn_default

# Predictions and accuracies on test data
# Based on random forest classifier
pred_labels <- predict(rf_default, new_data_test)
predictions_rf <- caret::confusionMatrix(table(pred_labels,new_data_test$label))
predictions_rf

# Based on knn classifier
pred_labels <- predict(knn_default, new_data_test)
predictions_knn <- caret::confusionMatrix(table(pred_labels,new_data_test$label))
predictions_knn
\end{verbatim}

Thus, it is easy to integrate results and objects that {\tt dbcsp} builds so that they can be integrated with other R packages and functions. This is interesting for more advanced users to perform  their own customized analysis.

\section{Conclusions}

In this work a new Distance-Based Common Spatial Pattern is introduced. It allows to perform the classical Common Spatial Pattern when the Euclidean distance between signals is considered, but it can be extended to the use of any other appropriate distance between signals as well. All of it is included in package {\tt dbcsp}. The package is easy to use for non specialized users but, in the sake of flexibility, more advanced analysis can be carried out combining the created object and obtained results with already well known R packages, as {\tt caret}, for instance.

\section*{Acknowledgements}
This research was partially supported: 
IR by The Spanish Ministry of Science, Innovation and Universities (FPU18/04737 predoctoral grant). II by the Spanish Ministerio de Economia y Competitividad (RTI2018-093337-B-I00; PID2019-106942RB-C31). CA by the Spanish Ministerio de Economia y Competitividad (RTI2018-093337-B-I00, RTI2018-100968-B-I00) and by Grant 2017SGR622 (GRBIO) from the Departament d'Economia i Coneixement de la Generalitat de Catalunya. BS  II by the Spanish Ministerio de Economia y Competitividad (RTI2018-093337-B-I00).

\section*{Author's contributions}
II and CA designed the study. IR and II wrote and debugged  the software. IR, II and CA  checked the software. II, CA, IR and BS wrote and reviewed the manuscript. All authors have read and approved the final manuscript.

\bibliography{Rodriguez.bib}      

\end{document}